\UseRawInputEncoding

\documentclass[12pt]{iopart}

\usepackage{iopams}  
\usepackage{graphicx} 
\usepackage{subfig}
\usepackage{multirow}
\usepackage{dsfont}
\usepackage{cite}
\usepackage{makecell}
\usepackage{xcolor}
\usepackage{colortbl}

\begin{document}

\title[Spectral statistics for the difference of two Wishart matrices]{Spectral statistics for the difference of two Wishart matrices}

\author{Santosh Kumar$^{*}$ \& S. Sai Charan }

\address{Department of Physics, Shiv Nadar University, Gautam Buddha Nagar, Uttar Pradesh -- 201314, India}
\ead{*skumar.physics@gmail.com}
\vspace{10pt}

\begin{abstract}
 In this work, we consider the weighted difference of two independent complex Wishart matrices and derive the joint probability density function of the corresponding eigenvalues in a finite-dimension scenario using two distinct approaches. The first derivation involves the use of unitary group integral, while the second one relies on applying the derivative principle. The latter relates the joint probability density of eigenvalues of a matrix drawn from a unitarily invariant ensemble to the joint probability density of its diagonal elements. Exact closed form expressions for an arbitrary order correlation function are also obtained and spectral densities are contrasted with Monte Carlo simulation results. Analytical results for moments as well as probabilities quantifying positivity aspects of the spectrum are also derived. Additionally, we provide a large-dimension asymptotic result for the spectral density using the Stieltjes transform approach for algebraic random matrices. Finally, we point out the relationship of these results with the corresponding results for difference of two random density matrices and obtain some explicit and closed form expressions for the spectral density and absolute mean.
\end{abstract}

%\maketitle

%%%%%%%%%% SECTION %%%%%%%%%%
\section{Introduction}

In recent years, there has been a surge in research activities related to composite random matrix ensembles which involve sums and products of matrices~\cite{AB2012,RE2008,AI2015,BJW2010,B2013,AIK2013,AKW2013,F2013,L2013,KZ2014,KS2014,ABKN2014,F2014a,F2014b,K2014,CKW2015,K2015a,K2015b,FL2016,ACK2016,WWKK2016,KKS2016,ARRS2016,AS2016a,AS2016b,WF2017,K2017,BSV2017,PKFD2017,PKF2017,MS2017,S2017,AC2018,FIL2018,FK2018,FIK2018,KPF2018,KPY2018,IN2018,IF2018,KR2019,FKK2019,KK2019,KFI2019,ACLS2019,K2019,Kumar2019,DF2020}. In this connection, in addition to the derivation of many crucial asymptotic and universal results, several integrable structures have been discovered which provide exact solutions to the spectral statistics of these ensembles. Aside from a mathematical interest, these composite ensembles naturally apply to a variety of problems in physics, engineering and related areas. For instance, sums and products of random matrices have been used in the context of multiple channel communication~\cite{AIK2013,AKW2013,K2017,PKF2017,PKFD2017,KPF2018,KPY2018}, quantum entanglement problem~\cite{L2013}, neutral network analyses~\cite{PB2017,G2020}, and random supergravity theories~\cite{MMW2012,PW2014,LMM2014}. Product of random matrices have also found applications in problems related to stochastic differential equations and Lyapunov exponents~\cite{B1954,FK1960,B1984,BL1985,CPV1993,F2013}, fixed point analysis for multi-layered complex systems~\cite{IF2018}, and Markov chains with random transition probabilities~\cite{IN2018}.

The difference between two covariance matrices appears in various measures used for their comparison~\cite{DKZ2009,PADS2014}. Since the covariance matrix resulting from a multivariate normal distribution is Wishart distributed, the difference between two Wishart matrices and the associated statistical quantities serves as a reference for their comparison purpose. Furthermore, the fixed (unit) trace variant of Wishart matrices model the Hilbert-Schmidt distributed density matrices~\cite{Forrester2010}, and has been extensively studied, see for example references~\cite{ZS2001,SZ2004,Nechita2007,KP2011,KSA2017,FK2019}. The difference between two such random density matrices plays a crucial role in quantum information theory in the context of distinguishability of quantum states \cite{H1969,M2007,AL2015,MZB2017,BSZW2016,PPZ2016,K2020}. In the present work, we consider weighted difference of two finite-dimensional independent complex Wishart matrices. This can be viewed as a multivariate generalization of weighted difference of two chi-square or gamma-distributed random variables. Unlike the sum of complex Wishart matrices, which is positive-definite Hermitian in nature, the difference is Hermitian but not positive definite. Therefore, it requires some additional precaution to obtain exact results pertaining to the spectral statistics of the difference of Wishart matrices. We obtain the joint probability density of the eigenvalues of the difference using two approaches. The first one involves the use of unitary group integral and leads to a joint eigenvalue density which falls under the category of polynomial ensembles of random matrix theory~\cite{KS2014,CKW2015,KKS2016,KK2019,FKK2019,KK2016,K2016,KR2019}. This category, in turn, belongs to that of Muttalib-Borodin biorthogonal ensembles~\cite{B1998,DF2008}. The second approach, on the other hand, relies on the application of the so called \emph{derivative principle} which gives a very interesting and powerful connection between the joint probability density of eigenvalues of a random matrix drawn from a unitarily invariant ensemble and that of its diagonal elements~\cite{CDKW2014,MZB2017,KZ2020}. This leads to the realization of a P\'olya ensemble, i.e., polynomial ensemble of a derivative type~\cite{KK2016,KR2019,FKK2019}. The polynomial-ensemble structure of the joint probability density enables us to write down the expressions for $r$-point correlation function in terms of determinants. We also consider large-dimension limit and provide asymptotic result for the spectral density using the Stieltjes transform approach for algebraic random matrices which relates to the computational free probability theory~\cite{RE2008}.

The presentation scheme in this paper is as follows. In section \ref{jpdev}, we derive two expressions for the joint probability density of eigenvalues of the weighted difference of two independent complex Wishart matrices at finite dimension. Section \ref{corrfunc} gives the corresponding exact results for the $r$-point correlation function. In section \ref{posmom} we obtain exact analytical results for probabilities quantifying positivity of the spectrum, and also spectral moments. In section \ref{SecAsymp}, asymptotic result for the spectral density is provided. In section~\ref{densmat}, we obtain some results for the spectral density and absolute mean for the difference of two random density matrices. We conclude in section~\ref{sumconc} with summary and outlook.

%%%%%%%%%% SECTION %%%%%%%%%%
\section{Joint eigenvalue density for difference of two complex Wishart matrices }
\label{jpdev}

We consider two independent $n$-dimensional matrices $W_1$ and $W_2$ from the complex Wishart distributions $\mathcal{CW}_n(n_1,\mathds{1}_n)$ and $\mathcal{CW}_n(n_2,\mathds{1}_n)$, respectively, as defined in equation~(\ref{PW}) below. Here, the parameter $n$ is the dimension of the Wishart matrices, and $n_1$ and $n_2$ are the numbers of degrees of freedom for the two matrices, respectively; see equation (\ref{ginib}). $\mathds{1}_n$ is the identity matrix of dimension $n$, which serves as the covariance matrix for both $W_1$ and $W_2$. The corresponding probability density functions (PDFs) are \cite{G1963},
\begin{equation}
\label{PW}
\mathcal{P}_j(W_j)=K_j \det(W_j)^{n_j-n}\exp(-\tr W_j)\Theta(W_j),~~j=1,2,
\end{equation}
where `$\det$' and `$\tr$' represent determinant and trace, respectively. The Heaviside theta function $\Theta(\cdot)$ with matrix argument imposes positive-definiteness condition. The normalization factor $K_j$ is given by \cite{G1963},
\begin{equation}
\label{Kj}
K_j= \left(\pi^{n(n-1)/2}\prod_{i=1}^n\Gamma(n_j-i+1)\right)^{-1}.
\end{equation}
The random matrix $W_j$ can be constructed as
\begin{equation}
\label{ginib}
W_j=G_jG_j^\dag,~~j=1,2,
\end{equation} 
where $G_j$ are ($n\times n_j$)-dimensional complex Ginibre random matrices governed by the PDFs 
\begin{equation}
\widetilde{\mathcal{P}}_j(G_j)=\pi^{-nn_j} e^{-\tr G_jG_j^\dag} .
\end{equation}
Our aim is to calculate the eigenvalue statistics of the random matrix given by the weighted difference,
\begin{equation}
\label{H1}
H=a_1 W_1-a_2 W_2,
\end{equation} 
where $a_1$ and $a_2$ are positive scalars.
It may be noted that the above is same as the difference $\widetilde{W}_1-\widetilde{W}_2$, where $\widetilde{W}_1$ and $\widetilde{W}_2$ are from the distributions $\mathcal{CW}_n(n_1,a_1^{-1}\mathds{1}_n)$ and $\mathcal{CW}_n(n_2,a_2^{-1}\mathds{1}_n)$. Evidently, $H$ is Hermitian but unlike $W_j$ it is not positive-definite in general.

%%%%%%%%%%
\subsection{Approach based on unitary group integral}

The PDF of $H$ can be calculated using
\begin{equation}
\mathcal{P}_H(H)=\int d[W_1]\int d[W_2]\delta\left(H-(a_1 W_1-a_2 W_2)\right)\mathcal{P}_1(W_1)\mathcal{P}_2(W_2),
\end{equation}
where the $\delta(\cdot)$ is the Dirac delta function with a matrix argument \cite{Zhang2016} and $d[W_j]$ represents the product of differentials of all the independent components in $W_j$, i.e., $d[W_j]=\prod_{i=1}^n dW_{j,ii}\prod_{1\le k<l\le n} \mathrm{Re}(W_{j,kl})\mathrm{Im}(W_{j,kl})$. The integral over $W_1$ can readily be done, leaving us with
\begin{eqnarray}
\nonumber
\mathcal{P}_H(H)&=&a_1^{-n^2}\int d[W_2]\mathcal{P}_1(a_1^{-1}H+a_1^{-1}a_2W_2)\mathcal{P}_2(W_2),\\
\nonumber
&=&K_1K_2\,a_1^{n(n-n_1)-n^2}e^{-a_1^{-1}\tr H}\int d[W_2]\det(W_2)^{n_2-n}e^{-\tr(1+a_1^{-1}a_2)W_2}\\
&&\times \det(H+a_2W_2)^{n_1-n}\Theta(H+a_2 W_2)\Theta(W_2).
\end{eqnarray}
We used here $\Theta(a_1^{-1}H+a_1^{-1}a_2W_2)=\Theta(H+a_2W_2)$, since $a_1>0$.
Now, we consider the eigenvalue decomposition, $W_2=\mathbb{U}^\dag \Omega\mathbb{U}$, where $\mathbb{U}$ is a unitary matrix and $\Omega$ is a diagonal matrix containing the eigenvalues $\{\omega_i\}$ of $W_2$. Also, let $\{\lambda_i\}$ be the eigenvalues of $H$. The integral over $W_2$ can then be transformed into the integrals over the eigenvalues and the unitary matrix. Noting that $\Theta(W_2)$ is equivalent to $\omega_i>0$ for $i=1,...n$, we obtain
\begin{eqnarray}
\nonumber
\fl
&&\mathcal{P}(\{\lambda_i\})
=K_1K_2\mathcal{V}^2\, a_1^{-nn_1}\Delta^2(\{\lambda_i\})\prod_{l=1}^n e^{-\lambda_l/a_1}\int_0^\infty d\omega_1\cdots \int_0^\infty d\omega_n \Delta^2(\{\omega_i\}) \\
\fl
&\times &\prod_{j=1}^n \omega_j^{n_2-n} e^{-(1+a_1^{-1}a_2)\omega_j}\int d\mu(\mathbb{U})\det(H+a_2\mathbb{U}^\dag \Omega \mathbb{U})^{n_1-n}\Theta(H+a_2\mathbb{U}^\dag \Omega \mathbb{U}).
\end{eqnarray}
Here, $\Delta(\{\omega_i\})=\prod_{j<k}(\omega_k-\omega_j)$ is the Vandermonde determinant and $\mathcal{V}=\pi^{n(n-1)/2}/\prod_{l=1}^n \Gamma(l+1)$ is the volume of the coset $\mathbb{U}(n)/[\mathbb{U}^n(1)\times\mathbb{S}(n)]$, with $\mathbb{U}(n)$ being unitary group comprising $n$-dimensional unitary matrices and $\mathbb{S}(n)$ is the permutation group of $n$ elements~\cite{ACK2016}. The integral over the above described coset can be performed using the result given in reference~\cite{KKS2016}. We obtain
\begin{eqnarray}
\fl
\mathcal{P}(\{\lambda_i\})
=K_1K_2\mathcal{V}^2\left(-\frac{1}{a_2}\right)^{n(n-1)/2} a_1^{-n n_1} \prod_{s=0}^{n-1}\left(n_1-1 \atop s\right)^{-1}\Delta(\{\lambda_i\})\prod_{l=1}^n e^{-\lambda_l/a_1} \\
\nonumber
\fl
\times\int_0^\infty d\omega_1\cdots \int_0^\infty d\omega_n \Delta(\{\omega_i\})\prod_{j=1}^n \omega_j^{n_2-n} e^{-(1+a_1^{-1}a_2)\omega_j}\det[(\lambda_j+a_2\omega_k)^{n_1-1}\theta(\lambda_j+a_2 \omega_k)]_{j,k=1}^n,
\end{eqnarray}
where $({a\atop b})=\Gamma(a+1)/(\Gamma(b+1)\Gamma(a-b+1))$ is the Binomial coefficient, and $\theta(\cdot)$ represents Heaviside theta function with a scalar argument.
The integrals over the eigenvalues $\omega_j$ can be performed using Andr\'eief's integral formula~\cite{Andreief1883}, which gives us the following after a little rearrangement of terms,
\begin{eqnarray}
\label{jpd1}
\mathcal{P}(\{\lambda_i\})
=C
 \Delta(\{\lambda_i\}) \det[f_j(\lambda_k)]_{j,k=1}^n.
\end{eqnarray}
Here, $f_j(\lambda)$ is a piecewise continuous function,
\begin{eqnarray}
\nonumber
&f_j(\lambda)=a_2^{j+n_2-n}e^{-\lambda/a_1}\int_0^\infty d\omega\,\omega^{n_2-n+j-1}e^{-(1+a_2/a_1)\omega}(\lambda+a_2\omega)^{n_1-1}\theta(\lambda+a_2\omega)\\
&= \cases{f_j^{(-)}(\lambda), & $\lambda<0$,\\ 
f_j^{(0)}(\lambda), & $\lambda=0$,\\
f_j^{(+)}(\lambda), & $\lambda>0$.}
\end{eqnarray}
with
\begin{eqnarray}
\fl 
f_j^{(-)}(\lambda)=\Gamma(n_1)(-\lambda)^{j+n_1+n_2-n-1}e^{\lambda/a_2}\,U\Big(n_1,j+n_1+n_2-n,-\Big(\frac{1}{a_1}+\frac{1}{a_2}\Big)\lambda\Bigg),\\
\fl 
f_j^{(0)}(\lambda)=\left(\frac{a_1a_2}{a_1+a_2}\right)^{j+n_1+n_2-n-1}\Gamma(j+n_1+n_2-n-1),\\
\fl \nonumber
f_j^{(+)}(\lambda)=\Gamma(j+n_2-n)\lambda^{j+n_1+n_2-n-1}e^{-\lambda/a_1}\,U\Big(j+n_2-n,j+n_1+n_2-n,\Bigg(\frac{1}{a_1}+\frac{1}{a_2}\Big)\lambda\Bigg).\\
\end{eqnarray}
%\begin{eqnarray}
%\nonumber
%\fl
%&f_j(\lambda)=a_2^{j+n_2-n}\int_0^\infty d\omega\,\omega^{n_2-n+j-1}e^{-(1+a_2/a_1)\omega}(\lambda_k+a_2\omega)^{n_1-1}\theta(\lambda_k+a_2\omega)\\
%\nonumber
%\fl
%&=\Gamma(j+n_2-n)\lambda^{j+n_1+n_2-n-1}e^{-\lambda/a_1}\,U\Big(j+n_2-n,j+n_1+n_2-n,\Bigg(\frac{1}{a_1}+\frac{1}{a_2}\Big)\lambda\Bigg)\theta(\lambda)\\
%\fl
%&+\Gamma(n_1)(-\lambda)^{j+n_1+n_2-n-1}e^{\lambda/a_2}\,U\Big(n_1,j+n_1+n_2-n,-\Big(\frac{1}{a_1}+\frac{1}{a_2}\Big)\lambda\Bigg)\theta(-\lambda)
%\end{eqnarray}
%It can be shown that, for $\lambda\to0$, both $\Gamma(j+n_2-n)\lambda^{j+n_1+n_2-n-1}e^{-\lambda/a_1}\,U\Big(j+n_2-n,j+n_1+n_2-n,\Bigg(\frac{1}{a_1}+\frac{1}{a_2}\Big)\lambda\Bigg)$
In the above expressions, $U(a,b,z)$ is the confluent hypergeometric function of the second kind (Tricomi function)~\cite{AS2012}.
The normalization factor is
\begin{eqnarray}
\label{C1}
\nonumber
C&=&(-1)^{n(n-1)/2}n! \,a_1^{-nn_1} a_2^{-nn_2}K_1K_2\mathcal{V}^2 \prod_{i=0}^{n-1}\left(n_1-1 \atop i\right)^{-1}\\
&=&\frac{(-1)^{n(n-1)/2}}{n! \,a_1^{nn_1}a_2^{nn_2}\prod_{j=1}^n\left[ \Gamma^2(j)\Gamma(n_1-j+1)\Gamma(n_2-j+1)\left(n_1-1 \atop j-1\right)\right]}.
\end{eqnarray}
On the other hand, application of Andreief's integral formula on~(\ref{jpd1}) yields an equivalent determinant based expression for $C$, given by
\begin{equation}
\label{C2}
C=(n! \det[h_{j,k}]_{j,k=1}^n)^{-1},
\end{equation}
where
\begin{eqnarray}
\label{hjk}
h_{jk}=\int_{-\infty}^\infty d\lambda\,\lambda^{k-1}f_j(\lambda)=h^{(-)}_{jk}+h^{(+)}_{jk}, 
\end{eqnarray}
with
\begin{eqnarray}
\label{hm}
 \nonumber
h_{jk}^{(-)}&=\int_{-\infty}^0 d\lambda\,\lambda^{k-1}f_j^{(-)}(\lambda)\\
\nonumber
&=(-1)^{k-1}\left(\frac{a_1a_2}{a_1+a_2}\right)^{j+k+n_1+n_2-n-1}(k+n_1)_{j+n_2-n-1}\,\Gamma(k)\Gamma(n_1)\\
&~~\times\,_2F_1\Big(k,j+k+n_1+n_2-n-1,k+n_1,\frac{a_2}{a_1+a_2}\Big),
\end{eqnarray}
\begin{eqnarray}
\label{hp}
 \nonumber
h_{jk}^{(+)}&=\int_0^\infty d\lambda\,\lambda^{k-1}f_j^{(+)}(\lambda)\\
 \nonumber
&=\left(\frac{a_1a_2}{a_1+a_2}\right)^{j+k+n_1+n_2-n-1}(j+k+n_2-n)_{n_1-1}\Gamma(k)\Gamma(j+n_2-n)\\
&~~\times\,\,_2F_1\Big(k,j+k+n_1+n_2-n-1,j+k+n_2-n,\frac{a_1}{a_1+a_2}\Big).
\end{eqnarray}
Here, $(a)_b=\Gamma(a+b)/\Gamma(a)$ and $_2F_1(a,b;c;z)$ are Pochhammer symbol and Gauss hypergeometric function~\cite{AS2012}, respectively.

%%%%%%%%%%
\subsection{Approach based on the derivative principle}
\label{SecDP}

\begin{figure}
\centering
\includegraphics[width=0.95\textwidth]{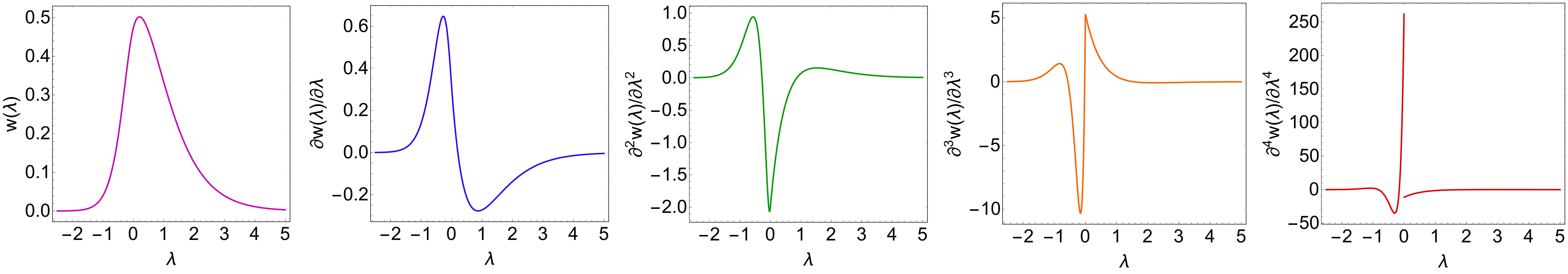}
\caption{Plots of $w(\lambda)$ as in equation (\ref{wHmm}) and its derivatives up to fourth order for $n_1=2,n_2=3,a_1=2/3,a_2=1/5$. We can see that the left and right derivatives do not agree at $\lambda=0$ in the last plot.}
\label{fig_diff}
\end{figure}

We now obtain an alternative expression for the joint PDF of eigenvalues of $H$ with the aid of the derivative principle, which asserts the following very interesting and powerful relationship~\cite{CDKW2014,MZB2017,KZ2020},
\begin{equation}
\label{dvpr}
\mathcal{P}(\{\lambda_i\}) =\frac{1}{\prod_{j=1}^n j!} \Delta(\{\lambda_i\})\Delta\left(\left\{-\frac{\partial}{\partial\lambda_i}\right\}\right)\mathcal{D}(\{\lambda_i\}).
\end{equation}
Here $\mathcal{D}(\{\cdot\})$ is the joint PDF of the diagonal elements of $H$. From equation~(\ref{ginib}), it is clear that the diagonal elements $W_{j,\mu\mu}=\sum_{\nu=1}^{n_j} |G_{\mu\nu}|^2$ are independent and chi-square distributed (or gamma distributed) with the PDF,
\begin{equation}
p_j(W_{j,\mu\mu})=\frac{1}{\Gamma(n_j)}W_{j,\mu\mu}^{n_j-1}e^{-W_{j,\mu\mu}}\theta(W_{j,\mu\mu}).
\end{equation}
Therefore, the diagonal elements of $H$ are weighted difference of two such independent random variables, viz., $H_{\mu\mu}=a_1W_{1,\mu\mu}-a_2W_{2,\mu\mu}$. The PDF of $H_{\mu\mu}$ can be obtained by the characteristic function method. For example, from Ref.~\cite{JB1970}, we have
\begin{eqnarray}
\label{wH}
\nonumber
\Phi(\kappa)=\int_0^\infty  dW_{1,\mu\mu} \int_0^\infty dW_{2,\mu\mu}e^{i\kappa(a_1W_{1,\mu\mu}-a_2W_{2,\mu\mu})}p_1(W_{1,\mu\mu})p_2(W_{2,\mu\mu})\\
\nonumber
=(1-i a_1 \kappa)^{-n_1}(1+i a_2 \kappa)^{-n_2}\\
\nonumber
=\sum_{\nu=1}^{n_1}\left({n_1+n_2-\nu-1}\atop{n_2-1}\right)\left(\frac{a_1}{a_1+a_2}\right)^{n_2}\left(\frac{a_2}{a_1+a_2}\right)^{n_1-\nu}(1-ia_1\kappa)^{-\nu}\\
+\sum_{\nu=1}^{n_2}\left({n_1+n_2-\nu-1}\atop{n_1-1}\right)\left(\frac{a_2}{a_1+a_2}\right)^{n_1}\left(\frac{a_1}{a_1+a_2}\right)^{n_2-\nu}(1+ia_2\kappa)^{-\nu}.
\end{eqnarray}
Performing the inverse Fourier transform, yields the PDF for $H_{\mu\mu}$,
\begin{eqnarray}
\label{wHmm}
\fl w(H_{\mu\mu})=\frac{1}{2\pi}\int_{-\infty}^\infty d\kappa\,e^{-i\kappa H_{\mu\mu}}\Phi(\kappa)
= \cases{w^{(-)}(H_{\mu\mu}), & $H_{\mu\mu}<0$,\\ 
w^{(0)}(H_{\mu\mu}), & $H_{\mu\mu}=0$,\\
w^{(+)}(H_{\mu\mu}), & $H_{\mu\mu}>0$,}
\end{eqnarray}
where
%\nonumber
%\fl=\sum_{\nu=1}^{n_1}\left({n_1+n_2-\nu-1}\atop{n_2-1}\right)\frac{a_1^{n_2-\nu}a_2^{n_1-\nu}}{(a_1+a_2)^{n_1+n_2-\nu}\Gamma(\nu)}H_{\mu\mu}^{\nu-1}e^{-H_{\mu\mu}/a_1}\Theta(H_{\mu\mu})\\
%\fl+\sum_{\nu=1}^{n_2}\left({n_1+n_2-\nu-1}\atop{n_1-1}\right)\frac{a_1^{n_2-\nu}a_2^{n_1-\nu}}{(a_1+a_2)^{n_1+n_2-\nu}\Gamma(\nu)}(-H_{\mu\mu})^{\nu-1}e^{H_{\mu\mu}/a_2}\Theta(-H_{\mu\mu}).
\begin{eqnarray}
\label{wm}
w^{(-)}(u)=\sum_{\nu=1}^{n_2}\left({n_1+n_2-\nu-1}\atop{n_1-1}\right)\frac{a_1^{n_2-\nu}a_2^{n_1-\nu}}{(a_1+a_2)^{n_1+n_2-\nu}\Gamma(\nu)}(-u)^{\nu-1}e^{u/a_2},\\
\label{w0}
w^{(0)}(u)=\frac{\Gamma(n_1+n_2-1)}{\Gamma(n_1)\Gamma(n_2)}\frac{a_1^{n_2-1}a_2^{n_1-1}}{(a_1+a_2)^{n_1+n_2-1}},\\
\label{wp}
 w^{(+)}(u)=\sum_{\nu=1}^{n_1}\left({n_1+n_2-\nu-1}\atop{n_2-1}\right)\frac{a_1^{n_2-\nu}a_2^{n_1-\nu}}{(a_1+a_2)^{n_1+n_2-\nu}\Gamma(\nu)}u^{\nu-1}e^{-u/a_1}.
\end{eqnarray}
The series in $w^{(-)}(u)$ and $w^{(+)}(u)$ above may be summed up in terms of the confluent hypergeometric function of the first kind (Kummer function)~\cite{AS2012}, $_1F_1(a;b;z)$, to yield
\begin{eqnarray}
\nonumber
\fl w^{(-)}(u)=\frac{\Gamma(n_1+n_2-1)}{\Gamma(n_1)\Gamma(n_2)}\frac{a_1^{n_2-1}a_2^{n_1-1}}{(a_1+a_2)^{n_1+n_2-1}}e^{u/a_2}\,_1F_1\left(1-n_2;2-n_1-n_2;-\Big(\frac{1}{a_1}+\frac{1}{a_2}\Big)u\right),\\
~\\
\nonumber
\fl w^{(+)}(u)=\frac{\Gamma(n_1+n_2-1)}{\Gamma(n_1)\Gamma(n_2)}\frac{a_1^{n_2-1}a_2^{n_1-1}}{(a_1+a_2)^{n_1+n_2-1}}e^{-u/a_1}\,_1F_1\left(1-n_1;2-n_1-n_2;\Big(\frac{1}{a_1}+\frac{1}{a_2}\Big)u\right).\\
\end{eqnarray}
Now, since $H_{\mu\mu}$ are independent, their joint PDF is given by $\mathcal{D}(H_{11},...,H_{nn})=\prod_{\mu=1}^n w(H_{\mu\mu})$. Therefore, from equation (\ref{dvpr}), we obtain the desired joint PDF of eigenvalues of $H$ as,
\begin{equation}
\mathcal{P}(\{\lambda_i\}) =\frac{1}{\prod_{j=1}^n j!} \Delta(\{\lambda_i\})\Delta\left(\left\{-\frac{\partial}{\partial\lambda_i}\right\}\right)\prod_{l=1}^n w(\lambda_{l}).
\end{equation}
This falls under the class of P\'olya ensemble of random matrices~\cite{KK2016,KR2019,FKK2019}. Next, we observe that $\Delta\left(\left\{-\frac{\partial}{\partial\lambda_i}\right\}\right)=(-1)^{n(n-1)/2}\det\left[\frac{\partial^{j-1}}{\partial \lambda_k^{j-1}}\right]_{j,k=1}^n$ acts on the product involving $w(\lambda_{l})$. Clearly, it can be recast in the form shown below.
\begin{eqnarray}
\label{jpd2}
\nonumber
\mathcal{P}(\{\lambda_i\}) &=&\frac{(-1)^{n(n-1)/2}}{\prod_{j=1}^n j!} \Delta(\{\lambda_i\})\det\left[\frac{\partial^{j-1}w(\lambda_{k})}{\partial \lambda_k^{j-1}}\right]_{j,k=1}^n\\
&=&\widetilde{C}\Delta(\{\lambda_i\})\det[\widetilde{f}_j(\lambda_k)]_{j,k=1}^n,
\end{eqnarray}
where 
\begin{equation}
\label{Ct1}
\widetilde{C}=\frac{(-1)^{n(n-1)/2}}{\prod_{j=1}^n \Gamma(j+1)},
\end{equation}
and $\widetilde{f}_j(\lambda)=\partial^{j-1}w(\lambda)/\partial \lambda^{j-1}$. Remarkably, $w(\lambda)$ is sufficiently smooth such that the derivatives $\partial^{j-1}w(\lambda)/\partial \lambda^{j-1}$ exist for $j=2,...,n$. In fact, as demonstrated in the~\ref{AppDiff}, the derivatives exist up to $j=n_1+n_2-1$ and hence $w(\lambda)$ is an $(n_1+n_2-2)$-times differentiable function. An example is shown in figure (\ref{fig_diff}), where we consider $n_1=2,n_2=3,a_1=2/3,a_2=1/5$ and plot $w(\lambda)$ and its derivatives. Now, as also demonstrated in the~\ref{AppDiff}, we find that
\begin{eqnarray}
\widetilde{f}_j(\lambda)
= \cases{\widetilde{f}_j^{(-)}(\lambda), & $\lambda<0$,\\ 
\widetilde{f}_j^{(0)}(\lambda), & $\lambda=0$,\\
\widetilde{f}_j^{(+)}(\lambda), & $\lambda>0$,}
\end{eqnarray}
%\begin{eqnarray}
%\nonumber
%\fl \widetilde{f}_j(\lambda)=\sum_{\nu=1}^{n_1}\left({n_1+n_2-\nu-1}\atop{n_2-1}\right)\frac{a_1^{n_2-\nu}a_2^{n_1-\nu}\Gamma(j)}{(a_1+a_2)^{n_1+n_2-\nu}\Gamma(\nu)}\lambda^{\nu-j}e^{-\lambda/a_1}L_{j-1}^{(\nu-j)}(\lambda/a_1)\Theta(\lambda)\\
%\nonumber
%\fl+\sum_{\nu=1}^{n_2}\left({n_1+n_2-\nu-1}\atop{n_1-1}\right)\frac{a_1^{n_2-\nu}a_2^{n_1-\nu}\Gamma(j)}{(a_1+a_2)^{n_1+n_2-\nu}\Gamma(\nu)}(-1)^{j-1}(-\lambda)^{\nu-j}e^{\lambda/a_2}L_{j-1}^{(\nu-j)}(-\lambda/a_2)\Theta(-\lambda),\\
%\end{eqnarray}
%with
%\begin{eqnarray}
%\fl\nonumber
%\widetilde{f}_j^{(-)}(\lambda)=\sum_{\nu=1}^{n_2}\left({n_1+n_2-\nu-1}\atop{n_1-1}\right)\frac{a_1^{n_2-\nu}a_2^{n_1-\nu}(-1)^{\nu-1}\Gamma(j)}{(a_1+a_2)^{n_1+n_2-\nu}\Gamma(\nu)}\lambda^{\nu-j}e^{\lambda/a_2}L_{j-1}^{(\nu-j)}\Big(-\frac{\lambda}{a_2}\Big),\\
%\fl\nonumber
%\widetilde{f}_j^{(0)}(\lambda)=\frac{\Gamma(n_1+n_2-1)}{\Gamma(n_1)\Gamma(n_2)}\frac{a_1^{n_2-1}a_2^{n_1-j}}{(a_1+a_2)^{n_1+n_2-1}}\,_2F_1\Big(1-j,1-n_2;2-n_1-n_2;\frac{a_1+a_2}{a_1}\Big)\\
%\nonumber
%\fl
%~~~~~~~~=\frac{\Gamma(n_1+n_2-1)}{\Gamma(n_1)\Gamma(n_2)}\frac{(-1)^{j-1}a_1^{n_2-j}a_2^{n_1-1}}{(a_1+a_2)^{n_1+n_2-1}}\,_2F_1\Big(1-j,1-n_1;2-n_1-n_2;\frac{a_1+a_2}%{a_2}\Big)\\
%\fl\widetilde{f}_j^{(+)}(\lambda)=\sum_{\nu=1}^{n_1}\left({n_1+n_2-\nu-1}\atop{n_2-1}\right)\frac{a_1^{n_2-\nu}a_2^{n_1-\nu}\Gamma(j)}{(a_1+a_2)^{n_1+n_2-\nu}\Gamma(\nu)}\lambda^{\nu-j}e^{-\lambda/a_1}L_{j-1}^{(\nu-j)}\Big(\frac{\lambda}{a_1}\Big).
%\end{eqnarray}
with
\begin{eqnarray}
\label{fm}\fl
\widetilde{f}_j^{(-)}(\lambda)=\sum_{\nu=1}^{n_2}\left({n_1+n_2-\nu-1}\atop{n_1-1}\right)\frac{(-1)^{\nu-1}a_1^{n_2-\nu}a_2^{n_1-j}}{(a_1+a_2)^{n_1+n_2-\nu}}e^{\lambda/a_2}L_{\nu-1}^{(j-\nu)}\Big(-\frac{\lambda}{a_2}\Big),\\
\label{f0}\fl\nonumber
\widetilde{f}_j^{(0)}(\lambda)=\frac{\Gamma(n_1+n_2-1)}{\Gamma(n_1)\Gamma(n_2)}\frac{a_1^{n_2-1}a_2^{n_1-j}}{(a_1+a_2)^{n_1+n_2-1}}\,_2F_1\Big(1-j,1-n_2;2-n_1-n_2;\frac{a_1+a_2}{a_1}\Big)\\
\fl
~~~~~~~~=\frac{\Gamma(n_1+n_2-1)}{\Gamma(n_1)\Gamma(n_2)}\frac{(-1)^{j-1}a_1^{n_2-j}a_2^{n_1-1}}{(a_1+a_2)^{n_1+n_2-1}}\,_2F_1\Big(1-j,1-n_1;2-n_1-n_2;\frac{a_1+a_2}{a_2}\Big),\\
\label{fp}
\fl\widetilde{f}_j^{(+)}(\lambda)=\sum_{\nu=1}^{n_1}\left({n_1+n_2-\nu-1}\atop{n_2-1}\right)\frac{(-1)^{\nu-j}a_1^{n_2-j}a_2^{n_1-\nu}}{(a_1+a_2)^{n_1+n_2-\nu}}e^{-\lambda/a_1}L_{\nu-1}^{(j-\nu)}\Big(\frac{\lambda}{a_1}\Big).
\end{eqnarray}
Here, $L_j^{(a)}(z)$ is the associated Laguerre polynomial \cite{Szego1975,AS2012}, which may be equivalently expressed in terms of confluent hypergeometric function using the relationship $L_j^{(a)}(z)=((a+1)_j/\Gamma(j+1))\,_1F_1(-j;a+1;z)$. Similar to equation~(\ref{C2}), the normalization factor can also be obtained using 
\begin{equation}
\label{Ct2}
\widetilde{C}=(n! \det[\widetilde{h}_{jk}]_{j,k=1}^n)^{-1},
\end{equation}
with
\begin{eqnarray}
\widetilde{h}_{jk}=\int_{-\infty}^\infty d\lambda\,\lambda^{k-1}\widetilde{f}_j(\lambda)=\widetilde{h}_{jk}^{(-)}+\widetilde{h}_{jk}^{(+)}.
\end{eqnarray}
Here,
\begin{eqnarray}
\fl\nonumber
\widetilde{h}_{jk}^{(-)}=\int_{-\infty}^0 d\lambda\,\lambda^{k-1}\widetilde{f}_j^{(-)}(\lambda)\\
\fl
=\sum_{\nu=1}^{n_2}\left({n_1+n_2-\nu-1}\atop{n_1-1}\right)\left({k-1}\atop{j-1}\right)\frac{(-1)^{k-1}a_1^{n_2-\nu}a_2^{n_1+k-j}}{(a_1+a_2)^{n_1+n_2-\nu}}\frac{\Gamma(j)\Gamma(\nu+k-j)}{\Gamma(\nu)},\\
\fl\nonumber
=\frac{(-1)^{k+1}\Gamma(k)\Gamma(n_1+n_2-1)}{\Gamma(n_1)\Gamma(n_2)}\frac{a_1^{n_2-1}a_2^{n_1-j+k}}{(a_1+a_2)^{n_1+n_2-1}}\,_2F_1\Big(1-j+k,1-n_2;2-n_1-n_2;\frac{a_1+a_2}{a_1}\Big),
\end{eqnarray}
and
\begin{eqnarray}
\fl\nonumber
\widetilde{h}_{jk}^{(+)}=\int_0^{\infty} d\lambda\,\lambda^{k-1}\widetilde{f}_j^{(+)}(\lambda)\\
\fl
=\sum_{\nu=1}^{n_1}\left({n_1+n_2-\nu-1}\atop{n_2-1}\right)\left({k-1}\atop{j-1}\right)\frac{(-1)^{j-1}a_1^{n_2+k-j}a_2^{n_1-\nu}}{(a_1+a_2)^{n_1+n_2-\nu}}\frac{\Gamma(j)\Gamma(\nu+k-j)}{\Gamma(\nu)}\\
\fl\nonumber
=\frac{(-1)^{j+1}\Gamma(k)\Gamma(n_1+n_2-1)}{\Gamma(n_1)\Gamma(n_2)} \frac{a_1^{n_2-j+k}a_2^{n_1-1}}{(a_1+a_2)^{n_1+n_2-1}}\,_2F_1\Big(1-j+k,1-n_1;2-n_1-n_2;\frac{a_1+a_2}{a_2}\Big).
\end{eqnarray}
It turns out that $\widetilde{h}_{jk}=\widetilde{h}_{jk}^{(-)}+\widetilde{h}_{jk}^{(+)}=0$ for $j>k$, and hence the normalization factor can be evaluated only using the diagonal elements, i.e., $\widetilde{C}=(n! \prod_{j=1}^n \widetilde{h}_{jj})^{-1}.$

We note that since equations (\ref{jpd1}) and (\ref{jpd2}) serve as the joint probability density of eigenvalues of $H$, they have to be equivalent. Indeed on evaluating them explicitly for a given set of parameters leads to identical expressions. We verified this in Mathematica using symbolic computation for several sets of parameter values. However, proving the equivalence for the general case seems difficult.
%%%%%%%%%%%% FIGURE %%%%%%%%%%%%%

%%%%%%%%%% SECTION %%%%%%%%%%
\section{Correlation functions}
\label{corrfunc}

\begin{figure}
\centering
\includegraphics[width=0.95\textwidth]{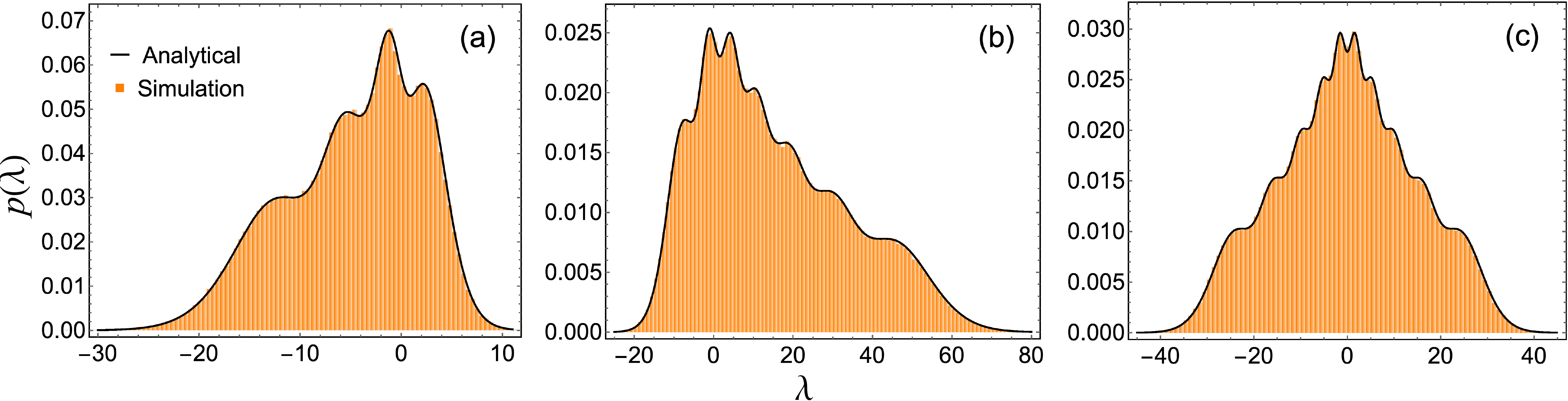}
\caption{Plots of the spectral density of weighted difference of Wisharts matrices, $H=a_1W_1-a_2W_2$. Comparison between analytical results (solid lines) and Monte Carlo simulations (histograms) for different sets of parameters: (a) $n = 4, n_1 = 5, n_2 = 7, a_1 = 2/3, a_2 = 8/7$ (b) $n= 7, n_1 = 11, n_2 = 7, a_1 = 2, a_2 = 1$ (c) $n = 10, n_1 = 11, n_2 = 11, a_1 = 1, a_2 = 1.$}
\label{fig_dens}
\end{figure}

The expressions for the joint probability density of eigenvalues of $H$ found in the preceding section are of polynomial-ensemble type~\cite{KS2014,K2016}, and therefore the $r$-point correlation function,
\begin{equation}
R_r(\lambda_1,...,\lambda_r)=\frac{n!}{(n-r)!}\int_{-\infty}^\infty d\lambda_{r+1}\cdots\int_{-\infty}^\infty d\lambda_{n} \mathcal{P}(\{\lambda_i\})
\end{equation}
can be written using the known results. A generalization of the Andr\'eief's integration formula~\cite{KG2010,K2015a} gives for the joint PDF given in equation~(\ref{jpd1}),
\begin{eqnarray}
\fl
\label{Rr}
R_r(\lambda_1,...,\lambda_r)=(-1)^r n! C\prod_{l=1}^r w(\lambda_l)\det\left[
\begin{array}{cc}
[0]_{j=1,...,r\atop k=1,...,r}  & [\lambda_j^{k-1}]_{j=1,...,r\atop k=1,...,n}   \\
\,[f_j(\lambda_k)]_{j=1,...,n \atop k=1,...,r} & [h_{jk}]_{j=1,...,n\atop k=1,...,n}
 \end{array}
\right].
\end{eqnarray}
Equivalently, we have~\cite{DF2008},
\begin{eqnarray}
\label{Rr1}
R_r(\lambda_1,...,\lambda_r)=\det[S(\lambda_j,\lambda_k)]_{j,k=1,...,r},
\end{eqnarray}
where the kernel $S(\lambda,\mu)$ is given by
\begin{eqnarray}
\label{Slm}
S(\lambda,\mu)=n! C\sum_{i=1}^n\lambda^{i-1} \det\Big[[h_{jk}]_{{j=1,...,n}\atop{k=1,...,i-1}}~[f_j(\mu)]_{{j=1,...,n}\atop{k=i}}~[h_{jk}]_{{j=1,...,n}\atop{k=i+1,...,n}}\Big].
\end{eqnarray}
Similarly, by replacing $C$, $f_{j}(\lambda_k)$ and $h_{jk}$ in the above equations by $\widetilde{C}$, $\widetilde{f}_{j}(\lambda_k)$ and $\widetilde{h}_{jk}$, we have the correlation function expression corresponding to equation~(\ref{jpd2}). We provide Mathematica~\cite{Mathematica} codes as ancillary files for obtaining the spectral density (the first order marginal density),
\begin{equation}
\label{plmbd}
p(\lambda)=S(\lambda,\lambda)/n.
\end{equation}
In figure \ref{fig_dens}, we show the spectral density based on the results obtained above for some combinations of the parameters. Monte Carlo simulation based results are also displayed using histograms. We can see a very good agreement in all cases.

%%%%%%%% FIGURE %%%%%%%%

%%%%%%%%%% SECTION %%%%%%%%%%

\section{Positivity of eigenvalues and spectral moments}
\label{posmom}

In this section, we compute some important quantities associated with the eigenvalues of the the matrix $H$. Since we are examining the eigenvalues of two positive-definite matrices, one is naturally interested in calculating the probability that the difference is also positive definite. Evidently, this equals the probability that all eigenvalues of $H$ are positive, which can be calculated as
\begin{equation}
\mathrm{P}^+=\int_0^\infty d\lambda_1\cdots \int_0^\infty d\lambda_n \mathcal{P}(\{\lambda_i\}).
\end{equation}
We can apply Andreief's integral formula to obtain the above from the equation (\ref{jpd1}),
\begin{equation}
\mathrm{P}^+=n! C \det[h^{(+)}_{jk}]_{j,k=1}^n=\frac{\det[h^{(+)}_{jk}]_{j,k=1}^n}{\det[h_{jk}]_{j,k=1}^n},
\end{equation}
where $C$, $h_{jk}$ and $h^{(+)}_{jk}$ are as in equations (\ref{C1}) or (\ref{C2}), (\ref{hjk}), and (\ref{hp}). Similarly, the probability that all eigenvalues of $H$ are negative is given by
\begin{eqnarray}
\nonumber
\mathrm{P}^-=\int_{-\infty}^0 d\lambda_1\cdots \int_{-\infty}^0 d\lambda_n \mathcal{P}(\{\lambda_i\})\\
=n! C \det[h^{(-)}_{jk}]_{j,k=1}^n=\frac{\det[h^{(-)}_{jk}]_{j,k=1}^n}{\det[h_{jk}]_{j,k=1}^n},
\end{eqnarray}
where $h^{(-)}_{jk}$ is given in (\ref{hm}). It should be noted that $\mathrm{P}^+ +\mathrm{P}^- \ne 1$, except when $n=1$.

Next, we examine the probability $\mathrm{p}^+$ of a generic eigenvalue of $H$ being positive, and therefore also get the information about the fraction of occurrence of positive eigenvalues on average. Moreover, it gives the probability that the spectral density has a positive support. We have,
\begin{equation}
\label{spp}
\mathrm{p}^+=\int_0^\infty d\lambda\, p(\lambda)=\frac{1}{n}\int_0^\infty d\lambda\, S(\lambda,\lambda).
\end{equation}
To evaluate the above, we use the expression of $S(\lambda,\lambda)$ from (\ref{Slm}), push the $\lambda^{i-1}$ appearing before the determinant in the $i$th column, and apply the $\lambda$-integral on this column. This gives us,
\begin{eqnarray}
\mathrm{p}^+=(n-1)! C\sum_{i=1}^n \det\Big[[h_{jk}]_{{j=1,...,n}\atop{k=1,...,i-1}}~[h_{jk}^{(+)}]_{{j=1,...,n}\atop{k=i}}~[h_{jk}]_{{j=1,...,n}\atop{k=i+1,...,n}}\Big].
\end{eqnarray}
In a similar fashion, the probability $\mathrm{p}^-$ of a generic eigenvalue of $H$ being negative is obtained as
\begin{eqnarray}
\mathrm{p}^-=(n-1)! C\sum_{i=1}^n \det\Big[[h_{jk}]_{{j=1,...,n}\atop{k=1,...,i-1}}~[h_{jk}^{(-)}]_{{j=1,...,n}\atop{k=i}}~[h_{jk}]_{{j=1,...,n}\atop{k=i+1,...,n}}\Big].
\end{eqnarray}
Here, we do have $\mathrm{p}^+ +\mathrm{p}^{-}=1$.

We now evaluate the spectral moment,
\begin{equation}
\langle\lambda^\gamma \rangle=\int_{-\infty}^\infty d\lambda\, \lambda^\gamma p(\lambda)=\frac{1}{n}\int_{-\infty}^\infty d\lambda\, \lambda^\gamma S(\lambda,\lambda).
\end{equation}
Similar to equation (\ref{spp}) above, this integral can also be evaluated to give us
\begin{eqnarray}
\langle\lambda^\gamma \rangle=(n-1)! C \sum_{i=1}^n \det\Big[[h_{jk}]_{{j=1,...,n}\atop{k=1,...,i-1}}~[h_{j,k+\gamma}]_{{j=1,...,n}\atop{k=i}}~[h_{jk}]_{{j=1,...,n}\atop{k=i+1,...,n}}\Big].
\end{eqnarray}
We can also calculate the absolute moment, i.e., the average of $|\lambda|^\gamma$. We have,
\begin{eqnarray}
\fl
\langle|\lambda|^\gamma \rangle=(n-1)! C \sum_{i=1}^n \det\Big[[h_{jk}]_{{j=1,...,n}\atop{k=1,...,i-1}}~[h_{j,k+\gamma}^{(+)}+(-1)^{-\gamma} h_{j,k+\gamma}^{(-)}]_{{j=1,...,n}\atop{k=i}}~[h_{jk}]_{{j=1,...,n}\atop{k=i+1,...,n}}\Big].
\end{eqnarray}
It is evident that when $\gamma$ is an even integer, we have $\langle|\lambda|^\gamma \rangle=\langle \lambda^\gamma \rangle$.

The above quantities can equivalently be written in terms of the $\widetilde{C}$, $\widetilde{h}^{(+)}_{jk}$, $\widetilde{h}^{(-)}_{jk}$ and $\widetilde{h}_{jk}$ which appear in section \ref{SecDP}.

%%%%%%%%%% SECTION %%%%%%%%%%
\section{Asymptotic result for the spectral density}
\label{SecAsymp}

The exact spectral density expressions obtained in the preceding section involves determinant. As such, it becomes difficult to obtain explicit symbolic expressions beyond $n=10$ or so. Therefore, in this section we obtain a large-dimension asymptotic spectral density expression analogous to Wigner's semicircle density or Mar\v{c}enko-Pastur density which can easily evaluate results even when $n$ is 1000 or more.
We consider the scaled Wishart matrices
\begin{equation}
\mathcal{W}_j=\frac{1}{n_j}W_j,~~j=1,2,
\end{equation}
so that the results in reference~\cite{RE2008} can directly be applied. We also define the parameters $c_j=n/n_j$. Clearly, we have $0<c_j\le 1$. Now, we consider large $n,n_1,n_2\gg1$ and proceed to obtain the asymptotic spectral density of 
\begin{equation}
\label{Hcal}
\mathcal{H}=\alpha_1\mathcal{W}_1-\alpha_2\mathcal{W}_2,
\end{equation}
for positive scalars $\alpha_1$ and $\alpha_2$. We note that the corresponding result for $H$ of equation (\ref{H1}) can be obtained by setting $\alpha_j=n_ja_j$. The above matrix model falls under the class of algebraic random matrices as defined in reference~\cite{RE2008} and therefore we can adopt the polynomial method approach developed therein. The key quantity in this context is the Stieltjes transform associated with the asymptotic spectral density $\hat{p}(\lambda)$,
\begin{equation}
s(z)=\int_{-\infty}^\infty \frac{\hat{p}(\lambda)}{\lambda-z}d\lambda,~~~z\in\mathbb{C}\setminus\mathbb{R}.
\end{equation}
The spectral density is recovered back using the Stieltjes-Perron inversion formula~\cite{A1965},
\begin{equation}
\label{SPinv}
\hat{p}(\lambda)=\frac{1}{\pi}\lim_{\epsilon\to 0^+} \mathrm{Im}[s(\lambda+i\epsilon)].
\end{equation}
We obtain the Stieltjes transform $s(z)$ of the asymptotic eigenvalue density for $\mathcal{H}$ using the methodology laid out in reference~\cite{RE2008}. It involves the stepwise construction of the desired composite matrix model from its constituents and calculation of the associated bivariate polynomials $L(s, z)$. By solving $L(s, z) = 0$ for $s$, in terms of $z$, one obtains the Stieltjes transform $s(z)$. The inversion formula then leads to the asymptotic spectral density. In Table I, we compile the implemented sequence of operations for constructing the matrix $\mathcal{H}$ from the matrices $\mathcal{W}_1$ and $\mathcal{W}_2$. From the table, we observe that $\mathcal{H} = A_4$ and therefore the Stieltjes transform for the associated eigenvalue density is obtained by solving $L_4(s, z) = 0$. The latter turns out to be a cubic equation in $s$,
\begin{equation}
\label{cubic}
g_3 z^3+g_2z^2+g_1z+g_0=0,
\end{equation}
where the $z$-dependent coefficients are given by
\begin{eqnarray}
\nonumber
g_3(z)=c_1 c_2 \alpha_1\alpha_2 z,\\
\nonumber
g_2(z)=(c_2\alpha_2-c_1\alpha_1)z+(c_1c_2-c_1-c_2)\alpha_1\alpha_2,\\
\nonumber
g_1(z)=-z+(1-c_1)\alpha_1-(1-c_2)\alpha_2,\\
g_0(z)=-1.
\end{eqnarray}
%%%% TABLE %%%%%
\begin{table}[!t]
\renewcommand{\arraystretch}{1.2}
\centering
\caption{Successive bivariate polynomials generated during the construction of $\mathcal{H}$ from $\mathcal{W}_1$ and $\mathcal{W}_2$.}
\begin{tabular}{ |l|l| }
  \hline
  {\bf Matrix} & {\bf Bivariate polynomial} \\ \hline
  $A_1=\mathds{1}_n$ & $L_1(s,z)=(1-z)s-1$ \\
  \hline
  $A_2=\mathcal{W}_2\cdot\mathds{1}_n$ & $L_2(s,z)=L_1\Big((1-c_2-c_2zs)s,\frac{z}{(1-c_2-c_2zs)}\Big)$ \\
    \hline
  $A_3=-\alpha_2A_2$ & $L_3(s,z)=L_2\Big(-\alpha_2s,-\frac{z}{\alpha_2}\Big)$ \\
   \hline
  $A_4=\alpha_1\mathcal{W}_1+A_3$ & $L_4(s,z)=L_3\Big(s,z-\frac{\alpha_1}{1+s\alpha_1c_1}\Big)$ \\
     \hline
\end{tabular}
\end{table}
%%% TABLE %%%
Once $s(z)$ is determined, the inversion formula (\ref{SPinv}) needs to be applied to arrive at $\hat{p}(\lambda)$. However, it is a nontrivial task to identify the ``correct" $s(z)$ out of the three roots. We know that, for a cubic equation, the roots can be expressed in terms of radicals and explicit results can be obtained, for example, using Cardano's method. After some effort, aided with Mathematica~\cite{Mathematica}, we are able to identify the correct solution and thereby obtain the asymptotic eigenvalue density. Let us define
\begin{equation}
\label{Glmbd}
G(\lambda)=\left[\frac{\eta(\lambda)+\sqrt{\eta^2(\lambda)-4\zeta^3(\lambda)}}{2}\right]^{1/3},
\end{equation}
with 
\begin{eqnarray}
\zeta(\lambda)=g_2^2(\lambda)-3g_1(\lambda)g_3(\lambda),\\
\eta(\lambda)=2g_2^3(\lambda)-9g_1(\lambda)g_2(\lambda)g_3(\lambda)+27g_0(\lambda)g_3^2(\lambda).
\end{eqnarray}
The asymptotic spectral density is then given by
\begin{equation}
\label{asymp}
\hat{p}(\lambda)=\frac{\mathrm{sign}(\lambda)}{2\sqrt{3}\,\pi g_3(\lambda)}\left(G(\lambda)-\frac{\zeta(\lambda)}{G(\lambda)}\right){\bf 1}_{[\lambda_{-},\lambda_{+}]},
\end{equation}
where ${\bf 1}_{[\lambda_{-},\lambda_{+}]}$ is the indicator function, meaning that the density is nonzero only in $[\lambda_{-},\lambda_{+}]$. The extremes $\lambda_{-}$ and $\lambda_{+}$ are determined from the two real roots of the quartic equation $[\eta^2(\lambda) - 4\zeta^3(\lambda)]/\lambda^2 = 0$. We note that $\eta^2(\lambda) - 4\zeta^3(\lambda)$ is a sixth degree polynomial in $\lambda$, but possesses $\lambda^2$ as a factor, and hence the division results in a quartic polynomial. We also observe that outside $[\lambda_{-},\lambda_{+}]$, $\eta^2(\lambda) - 4\zeta^3(\lambda)$ becomes negative, thereby rendering its square-root as complex. It is possible to obtain the roots of $[\eta^2(\lambda) - 4\zeta^3(\lambda)]/\lambda^2$ explicitly using Ferrari's method. However, it is known that these results are extremely lengthy and therefore it is far more convenient to simply locate the extremes by solving the quartic equation numerically. One important point to note is that the cube root in equation (\ref{Glmbd}) should be evaluated to a real value if its argument happens to be negative, e.g. $(-8)^{1/3} = -2$, and not one of the two other complex solutions.

As already mentioned, the asymptotic eigenvalue density for $H$ can be obtained by setting $\alpha_j=n_ja_j$ in the above expressions. We should remark that the above asymptotic result also applies to the difference of real Wishart matrices with probability densities proportional to $\det(W_j)^{(n_j-n-1)/2}\exp(-\frac{1}{2}\tr W_j)\Theta(W_j),~~j=1,2$; {\it cf.} equation (\ref{PW}). We provide Mathematica code for computing the above asymptotic density in the ancillary material.

\begin{figure}
\centering
\includegraphics[width=0.95\textwidth]{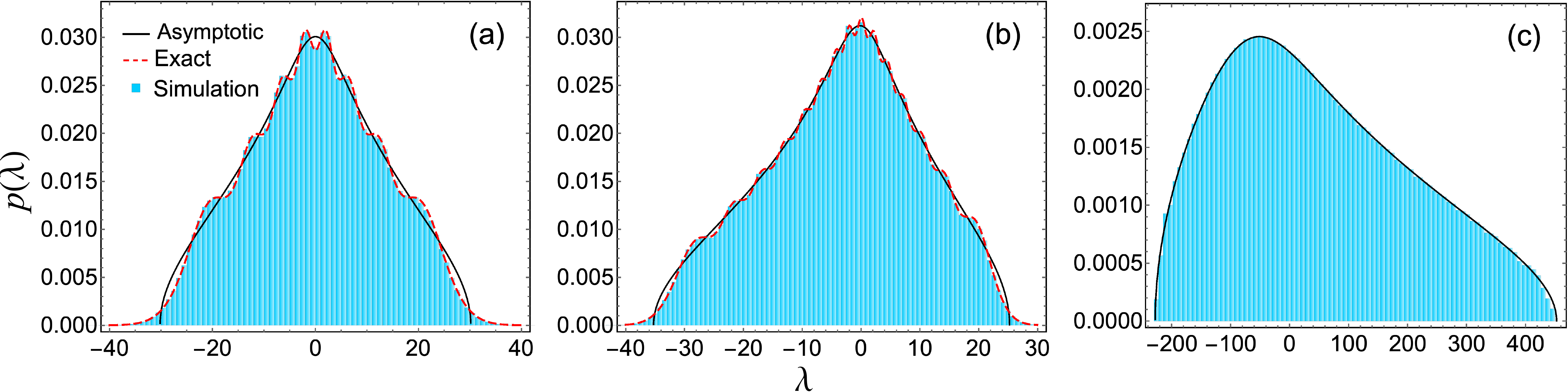}
\caption{Plots of the spectral density of weighted difference of Wisharts matrices, $H=a_1W_1-a_2W_2$. Comparison between asymptotic results (solid lines) and Monte Carlo simulations (histograms) for different sets of parameters: (a) $n = 8, n_1 = 11, n_2 = 11, a_1 = 1, a_2 = 1$ (b) $n= 15, n_1 = 18, n_2 = 20, a_1 = 1/2, a_2 = 3/5$ (c) $n = 50, n_1 = 100, n_2 = 200, a_1 = 2, a_2 = 3/4$. In plots (a) and (b) exact results for the spectral density are also shown with the aid of dashed lines.}
\label{fig_asymp}
\end{figure}

In figure~\ref{fig_asymp}, we show the comparison of asymptotic densities along with the Monte Carlo simulation results. We find that already for values of $n$ as small as 8, the asymptotic result performs fairly well. For large dimension, it is practically indistinguishable from the empirical density obtained from simulation, as can be seen in the $n=50$ case.
%%%%%%%% FIGURE %%%%%%%%
%%%%%%%% FIGURE %%%%%%%%

%%%%%%%%%% SECTION %%%%%%%%%%
\section{Difference of two random density matrices}
\label{densmat}

The difference of two random density matrices, the Helstrom matrix, plays a fundamental role in quantum information theory. Various distance measures involve this difference and, among other things, it is quite useful in distinguishing given quantum states~\cite{M2007,MZB2017,BSZW2016,PPZ2016,K2020}. Let us consider two random density matrices $\rho_1$ and $\rho_2$ of dimension $n$ distributed according to the Hilbert-Schmidt measure, viz.,
\begin{equation}
\mathcal{P}^{(F)}_j(\rho_j)=\Gamma(nn_j)K_j\det(\rho_j)^{n_j-n}\delta(\tr\rho_j-1)\Theta(\rho_j),
\end{equation}
where $K_j$ is as in equation (\ref{Kj}). In the context of random matrix theory, the above corresponds to fixed (unit) trace Wishart ensemble, and hence we have used the superscript `$(F)$' above. The above measure over the $n$-dimensional reduced state $\rho_1$ (and $\rho_2$) is induced by the operation of partial tracing over the other constituent of dimension $n_1$ (resp., $n_2$) from a random bipartite pure state living in an $nn_1$ (resp., $nn_2$) -dimensional Hilbert space. If a single bipartite system is considered, then $\rho_1$ and $\rho_2$ need to be independent.
%%%%%%FIGURE%%%%%%
\begin{figure}[!ht]
\centering
\includegraphics[width=0.95\textwidth]{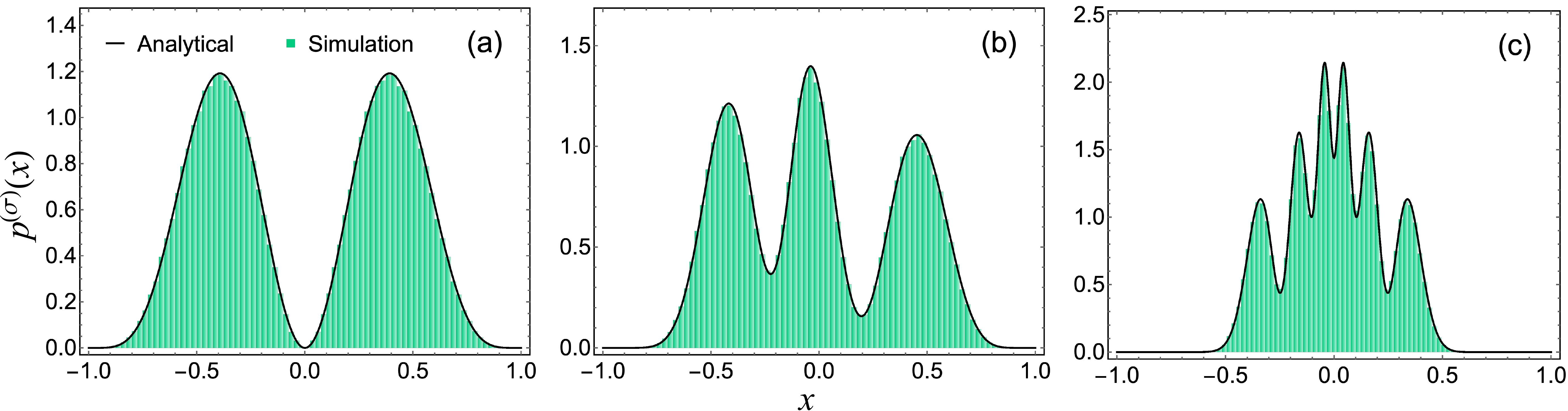}
\caption{Plots of the spectral density of difference of two random density matrices, $\sigma=\rho_1-\rho_2$. Comparison between analytical results and Monte Carlo simulations for different values of parameters: (a) $n = 2, n_1 = 3, n_2 = 4$ (b) $n= 3, n_1 = 3, n_2 = 5$ (c) $n = 6, n_1 = 6, n_2 = 6$.}
\label{fig_sigma}
\end{figure}
%%%%%%%% FIGURE %%%%%%%%
\begin{figure}
\centering
\includegraphics[width=0.95\textwidth]{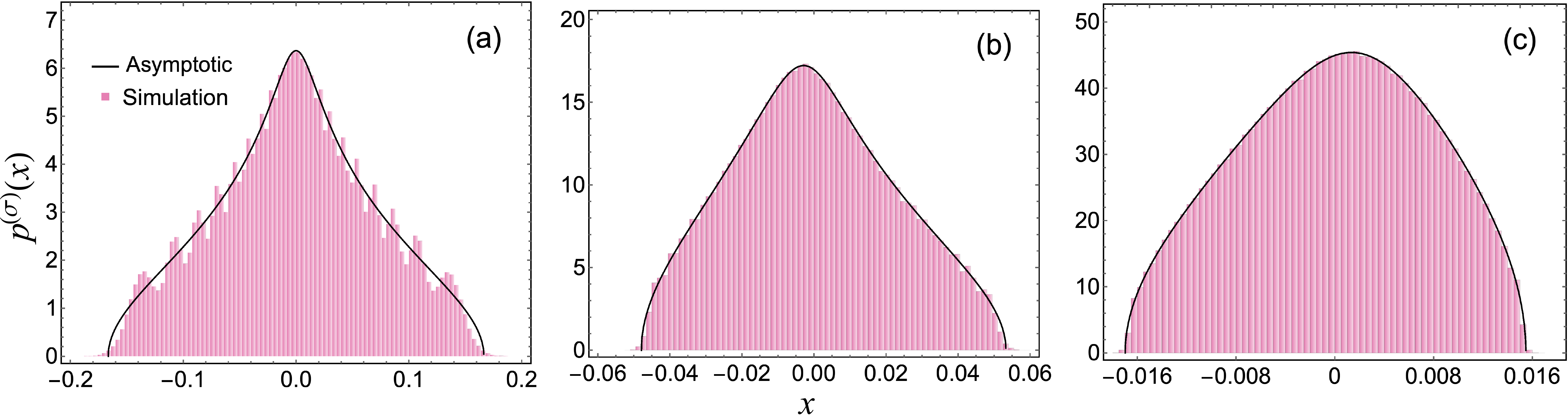}
\caption{Plots of the spectral density of difference of two random density matrices, $\sigma=\rho_1-\rho_2$. Comparison between large-dimension asymptotic results and Monte Carlo simulations for different choices of parameters: (a) $n = 20, n_1 = 20, n_2 = 20$ (b) $n= 50, n_1 = 70, n_2 = 90$ (c) $n = 100, n_1 = 400, n_2 = 300$.}
\label{dens_asymp}
\end{figure}
%%%%%%%%%%%%%%%%
The joint probability density of eigenvalues of the difference
\begin{equation}
\sigma=\rho_1-\rho_2,
\end{equation}
has been derived in reference~\cite{MZB2017} for $n_1=n_2$ case using the derivative principle by first obtaining the joint probability density of diagonal elements of $\sigma$ as a contour integral. However, due to its complicated structure, an explicit expression for the spectral density could be obtained for $n=2$ only. Fortunately, the asymptotic spectral density is known in the large dimension limit for $n_1=n_2$ case~\cite{MZB2017,PPZ2016}. In the following, we exploit the relationship of $\sigma$ with the difference $H$ of Wishart matrices to obtain explicit expressions of the spectral density for some low-dimension cases in the general scenario when $n_1$ and $n_2$ may differ. Furthermore, we provide asymptotic result for large dimension limit in this general case.

It can be shown that for the Helstrom matrix $\sigma$, the matrix probability density, joint probability density of eigenvalues and all correlation functions can be obtained from the corresponding expressions for the difference of Wishart matrices using a double Fourier-Laplace inversion. This is similar to the well known Fourier-Laplace relationship between a Wishart matrix and its fixed-trace variant. We focus here on the spectral density. Let use include the weights $a_1,a_2$ explicitly as arguments in spectral density expression for difference of Wishart matrices, i.e. use the notation $p(a_1,a_2,\lambda)$ instead of $p(\lambda)$ given in (\ref{plmbd}). Then, the desired spectral density for $\sigma$ is
\begin{equation}
\fl
p^{(\sigma)}(x)=\Gamma(nn_1)\Gamma(nn_2)\mathcal{L}^{-1}[s_1^{-nn_1}s_2^{-nn_2}p(s_1^{-1},s_2^{-1},x),\{s_1,s_2\},\{t_1,t_2\}]_{t_1=1=t_2}.
\end{equation} 
Likewise, the spectral moments in the present case can be related to those of the matrix $H$. 
However, due to complicated structure of the expressions involved, it is not feasible to obtain general results in the present case by performing the double Laplace inversion. Nonetheless, we may obtain explicit answers for small values of $n,n_1,n_2$ on a case-by-case basis. For $n=2$, there is an $x\to-x$ symmetry in $p^{(\sigma)}(x)$ for any $n_1,n_2$, i.e., the densities are symmetric about $x=0$. For $n>2$, if the spectral density for $n_2\ge n_1$ case is $p^{(\sigma)}(x)$, then for $n_2\le n_1$ it is given by $p^{(\sigma)}(-x)$. It should also be noted that the density vanishes outside $[-1,1]$.

 We compile the exact results for $p^{(\sigma)}(x)$ and the average $\langle |x|\rangle_{\sigma}=\int_{-1}^1 dx|x|\,p^{(\sigma)}(x)$,  for $2\le n\le n_1\le n_2\le 4$ in Table 2. The average of $|x|$ is related to the trace-distance between the density matrices $\rho_1$ and $\rho_2$ and gives the average optimal probability of distinguishing them~\cite{PPZ2016,MZB2017,H1969,M2007,AL2015}. Further exact results for $p^{(\sigma)}(x)$ and $\langle |x|\rangle_{\sigma}$ are provided for $2\le n\le n_1\le n_2\le 6$ in an accompanying Mathematica file along with plots of the spectral density. As far as the positivity of support $p^{+}$ is concerned for the difference of two density matrices, due to the above discussed symmetry of the spectral density, it is 1/2 for $n=2$ and arbitrary $n_1,n_2$ and for $n>2$, it is again 1/2 when $n_1=n_2$. For other combinations, one may attempt to evaluate $p^{+}$ on a case-by-case basis using the computed spectral density. The positivity aspect is useful for the study of entanglement criteria in quantum information theory, see for example the reference~\cite{BN2013} for the case of the partial transposition.

In figure \ref{fig_sigma}, we show the plots for spectral density of difference of two random density matrices for some example values of parameters. Monte Carlo simulation results are also shown and agree very well with the analytical results.

\begin{table}[h]
\renewcommand{\arraystretch}{1.5}
\centering
\caption{Spectral density $p^{(\sigma)}(x)$ for Helstrom matrix $\sigma=\rho_1-\rho_2$ and the corresponding absolute mean $\langle|x|\rangle_\sigma$ for various combinations of the dimensions $n,n_1$ and $n_2$. The notation $\{x\to -x\}$ means that the expression for $0\le x\le 1$ follows from that of $-1\le x\le 0$ by changing $x$ to $-x$. Outside $[-1,1]$, the density is zero.}
\begin{tabular}{ |c|c|c|c c| c|}
  \hline
  $n$ &$n_1$ & $n_2$ & $p^{(\sigma)}(x)$ & ~ & $\langle|x|\rangle_\sigma$ \\ \hline\hline
  2 & 2 & 2 & $\makecell{6x^2(1+x)^2(2-x), \\ \{x\to -x\}, }$ & $\makecell{-1\le x\le 0 \\ ~~0\le x\le 1 }$ & $\displaystyle\frac{18}{35}$\\
  \hline
  2 & 2 & 3 & $\makecell{12x^2(1+x)^3(1-3x+x^2), \\  \{x\to -x\},} $ & $\makecell{-1\le x\le 0 \\ ~~0\le x\le 1 }$ & $\displaystyle\frac{10}{21}$\\
  \hline
  2 & 2 & 4 & $\makecell{6x^2(1+x)^4(2-8x+20x^2-5x^3), \\  \{x\to -x\},} $ & $\makecell{-1\le x\le 0 \\ ~~0\le x\le 1  }$ & $\displaystyle\frac{5}{11}$\\
  \hline
  2 & 3 & 3 & $\makecell{\frac{30}{7}x^2(1+x)^4(4-16x+12x^2-3x^3), \\  \{x\to -x\},} $ & $\makecell{-1\le x\le 0 \\ ~~0\le x\le 1  }$ & $\displaystyle\frac{100}{231}$\\
  \hline
  2 & 3 & 4 & $\makecell{ 20x^2(1+x)^5(1-5x+9x^2-5x^3+x^4),\\  \{x\to -x\},} $ & $\makecell{-1\le x\le 0 \\ ~~0\le x\le 1  }$ & $\displaystyle\frac{175}{429}$ \\
  \hline
  2 & 4 & 4 & $\makecell{\frac{140}{33}x^2(1+x)^6(6-36x+82x^2-72x^3+30x^4-5x^5), \\  \{x\to -x\},} $ & $\makecell{-1\le x\le 0 \\ ~~0\le x\le 1  }$ & $\displaystyle\frac{490}{1287}$\\
  \hline
  3 & 3 & 3 & $\makecell{\frac{8}{143}(1+x)^6(24-144x-1064x^2+2058x^3+9772x^4\\-18158x^5+18088x^6-11753x^7+4494x^8-749x^9), \\  \{x\to -x\},} $ & $\makecell{~\\-1\le x\le 0 \\ ~~0\le x\le 1  }$ & $\displaystyle\frac{1184}{3315}$\\
  \hline
  3 & 3 & 4 & $\makecell{\frac{2}{663}(1+x)^8(440-4675x+3507x^2+182244x^3\\-113388x^4-1883250x^5+2377170x^6-1645656x^7\\+727788x^8-193351x^9+23495x^{10}),\\
  \frac{2}{663}(1-x)^7(440+1925x-17338x^2-42351x^3\\+164496x^4+519078x^5+826140x^6+925386x^7\\+773052x^8+458963x^9+159074x^{10}+23495x^{11})}$ & $\makecell{~\\~\\-1\le x\le 0 \\~\\~\\ ~~0\le x\le 1  }$ & $\displaystyle\frac{979}{2907}$\\
     \hline
 3 & 4 & 4 & $\makecell{\frac{220}{88179}(1+x)^9(572-5148 x-6831 x^2+198759 x^3\\-76527 x^4-2406294 x^5+4903878 x^6-5613012 x^7\\+4481748 x^8-2583459 x^9+1026651 x^{10}-249615 x^{11}\\+27735 x^{12}),\\  \{x\to -x\}, }$   & $\makecell{~\\~\\~\\-1\le x\le 0 \\ ~~0\le x\le 1  }$ & $\displaystyle\frac{48950}{156009}$\\
   \hline  
  4 & 4 & 4 & $\makecell{\frac{2}{646323}(1+x)^{12}(251940-3023280 x+79319110 x^2\\-807719640 x^3-3849356784 x^4+12770088968 x^5\\+23325866928 x^6-70508450649 x^7+97987112860 x^8\\-97689979023 x^9+77811833736 x^{10}-51349726064 x^{11}\\+28242904872 x^{12}-12756361800 x^{13}+4561977896 x^{14}\\-1207136637 x^{15}+208188708 x^{16}-17349059 x^{17}),\\  \{x\to -x\}, }$  & $\makecell{~\\~\\~\\~\\~\\-1\le x\le 0 \\ ~~0\le x\le 1  }$ & $\displaystyle\frac{1495}{5394}$ \\
 \hline
\end{tabular}
\end{table}

For asymptotic limit of large dimensions, viz., $1\ll n\le n_1\le n_2$, the result from the preceding section can be used to obtain the density here also. The average trace for both the matrices $\mathcal{W}_1$ and  $\mathcal{W}_2$ is $n$. Therefore, in the asymptotic limit, we consider $\alpha_1=\alpha_2=1/n$ in equation (\ref{Hcal}) and then the equation~(\ref{asymp}) captures the spectral density of difference of two random density matrices. We show some examples in figure \ref{dens_asymp}.

%\begin{table}
 % \renewcommand{\arraystretch}{1.5}
%\textcolor{blue}{
 % \caption{Mode Transition Times}
 % \centering
 % \begin{tabular}{|c|c|c|c|c|c|c|c|}
  %  \hline
  %  \multirow{2}{1cm}{~~~$n$} &    \multirow{2}{1cm}{~~$n_1$} &  \multirow{2}{1cm}{~~$n_2$}& \multicolumn{2}{c|}{$p^{+}$} & \multicolumn{2}{c|}{$\langle|x|\rangle_\sigma$}\\
    % \hline
    % \textbf{Inactive Modes} & \textbf{Description}\\
  %  \cline{4-7}
  %  & ~& ~ & Exact & Numerical Value & Exact & Numerical Value\\
    %\hhline{~--}
  %  \hline
   % 2 & 2 & 2 & 1024 & $1/2$ s & $\displaystyle\frac{18}{35}$ & g\\ \hline
  %  2 & 2 & 3 & 1024 & $1/2$ s & $\displaystyle\frac{10}{21}$ & g\\ \hline
  %  2 & 2 & 4 & 1024 & $1/2$ s & $\displaystyle\frac{5}{11}$ & g\\ \hline
  %  2 & 3 & 3 & 1024 & $1/2$ s & $\displaystyle\frac{100}{231}$ & g\\ \hline
  %  2 & 3 & 4 & 1024 & $1/2$ s & $\displaystyle\frac{175}{429}$ & g\\ \hline
   % 2 & 4 & 4 & 1024 & $1/2$ s & $\displaystyle\frac{490}{1287}$ & g\\ \hline
  %  3 & 3 & 3 & 1024 & $1/2$ s &  $\displaystyle\frac{1184}{3315}$ & g\\ \hline
  %  3 & 3 & 4 & 1024 & $~$ s & $\displaystyle\frac{979}{2907}$ & g\\ \hline
  %  3 & 4 & 4 & 1024 & $1/2$ s & $\displaystyle\frac{48950}{156009}$ & g\\ \hline
  %  4 & 4 & 4 & 1024 & $1/2$ s & $\displaystyle\frac{1495}{5394}$ & g\\ \hline
 % \end{tabular}
 % }
%\end{table}

%%%%%%%%%% SECTION %%%%%%%%%%
\section{Summary and Conclusion}
\label{sumconc}

In this work, we derived the joint probability density of eigenvalues of weighted difference of two Wishart matrices of finite dimension using two approaches. Corresponding exact expressions for the correlation functions were also provided. Spectral moments and quantities to assess positivity of the eigenvalues were also derived. Additionally, we derived a large-dimension asymptotic result for the spectral density. These results were then employed to obtain some exact finite-dimension results and also a large-dimension asymptotic result for the spectral density of difference of two random density matrices. We compared our analytical results with Monte Carlo simulations and found very good agreements. 

Our focus in the present work was on the global spectral statistics of the difference of two Wishart matrices and two random density matrices. It would be of much interest to investigate the corresponding extreme eigenvalues, as well as local aspects of the spectra which would reveal the universal features of these composite random matrix ensembles.

%%%%%%%%%% SECTION %%%%%%%%%%
%\section{Acknowledgments}

\pagebreak
%%%%%%%%%% Appendices %%%%%%%%%%
\appendix

\section{Differentiability of $w(x)$}
\label{AppDiff}

We prove here that the derivative $\partial^{j-1}w(\lambda)/\partial\lambda^{j-1}$ exists for $j=2,...,n_1+n_2-1$. It is clear from equation (\ref{wH}) that $w(\lambda)$ is analytic for both $\lambda< 0$ and $\lambda>0$, and the derivatives in these regions can be obtained readily. The expressions given in (\ref{fm}) and (\ref{fp}) follow from (\ref{wm}) and (\ref{wp}) by applying Rodrigues formula~\cite{Szego1975},
\begin{equation}
\label{rodrigues}
\frac{d^k(e^{-x} x^{k+a})}{dx^k}=\Gamma(k+1) e^{-x}x^aL_k^{(a)}(x),
\end{equation}
followed by the result~\cite{Szego1975},
\begin{equation}
\label{lagrel}
L_k^{(a)}(x)=(-x)^{-a}\frac{\Gamma(k+a+1)}{\Gamma(k+1)}L_{k+a}^{(-a)}(x),
\end{equation}
so that,
\begin{equation}
%\label{rodrigues}
\frac{d^k(e^{-x} x^{k+a})}{dx^k}=(-1)^{a}\Gamma(k+a+1)e^{-x}L_{k+a}^{(-a)}(x).
\end{equation}

We now demonstrate the differentiability of $w(\lambda)$ at $\lambda=0$ below. To this end, we show that
\begin{eqnarray}
\nonumber
\lim_{\epsilon\to0^{-}}\frac{1}{\epsilon^{j-1}}\sum_{\mu=0}^{j-1}(-1)^{\mu+j-1}\left({j-1 \atop \mu}\right)w(\mu\epsilon)\\
=\lim_{\epsilon\to0^{+}}\frac{1}{\epsilon^{j-1}}\sum_{\mu=0}^{j-1}(-1)^\mu\left({j-1 \atop \mu}\right)w((j-1-\mu)\epsilon),
\end{eqnarray}
i.e., the left and right derivatives are equal for $j=2,...,n_1+n_2-1$. Focussing on the left derivative, we have,
\begin{eqnarray}
\label{lrderiv}
\nonumber
\lim_{\epsilon\to0^{-}}\frac{1}{\epsilon^{j-1}}\sum_{\mu=0}^{j-1}(-1)^{\mu+j-1}\left({j-1 \atop \mu}\right)w(\mu\epsilon)\\
\nonumber
=\lim_{\epsilon\to0}\frac{(-1)^{j-1}}{\epsilon^{j-1}}\left[w^{(0)}(0)+\sum_{\mu=1}^{j-1}(-1)^\mu\left({j-1 \atop \mu}\right)w^{(-)}(\mu\epsilon)\right]\\
\label{dwm}
=\lim_{\epsilon\to0}\frac{(-1)^{j-1}}{\Gamma(j)}\sum_{\mu=1}^{j-1}(-1)^\mu\left({j-1 \atop \mu}\right)\frac{\partial^{j-1}w^{(-)}(\mu\epsilon)}{\partial \epsilon^{j-1}},
\end{eqnarray}
where we used L'Hospital's rule in the last step. Now, we use the expression for $w^{(-)}(\nu\epsilon)$ from equation (\ref{wm}) to obtain its $(j-1)$-th derivative with respect to $\epsilon$. We have
\begin{eqnarray}
\nonumber
\fl
\frac{\partial^{j-1}w^{(-)}(\mu\epsilon)}{\partial \epsilon^{j-1}}=\sum_{\nu=1}^{n_2}\left({n_1+n_2-\nu-1}\atop{n_1-1}\right)\frac{a_1^{n_2-\nu}a_2^{n_1-\nu}}{(a_1+a_2)^{n_1+n_2-\nu}\Gamma(\nu)}\frac{\partial^{j-1}}{\partial \epsilon^{j-1}}(-\mu\epsilon)^{\nu-1}e^{\mu\epsilon/a_2}\\
\nonumber
\fl
=\sum_{\nu=1}^{n_2}\left({n_1+n_2-\nu-1}\atop{n_1-1}\right)\frac{(-1)^{\nu-1}a_1^{n_2-\nu}a_2^{n_1-\nu}\Gamma(j)}{(a_1+a_2)^{n_1+n_2-\nu}\Gamma(\nu)}\mu^{j-1}(\mu\epsilon)^{\nu-j}e^{\mu\epsilon/a_2}L_{j-1}^{(\nu-j)}\Big(-\frac{\mu\epsilon}{a_2}\Big)\\
\fl
=\sum_{\nu=1}^{n_2}\left({n_1+n_2-\nu-1}\atop{n_1-1}\right)\frac{(-1)^{\nu-1}a_1^{n_2-\nu}a_2^{n_1-j}}{(a_1+a_2)^{n_1+n_2-\nu}}\mu^{j-1}e^{\mu\epsilon/a_2}L_{\nu-1}^{(j-\nu)}\Big(-\frac{\mu\epsilon}{a_2}\Big).
\end{eqnarray}
In the second step of the above calculation we used Rodrigues formula (\ref{rodrigues}) and in last step we used the relation (\ref{lagrel}). Now, the limit $\epsilon\to0$ can be conveniently taken by using the result $L_k^{(a)}(0)=\left({k+a\atop k}\right)$, giving us
\begin{eqnarray}
\nonumber
\fl
\lim_{\epsilon\to0}\frac{\partial^{j-1}w^{(-)}(\mu\epsilon)}{\partial \epsilon^{j-1}} =\sum_{\nu=1}^{n_2}\left({n_1+n_2-\nu-1}\atop{n_1-1}\right)\frac{(-1)^{\nu-1}a_1^{n_2-\nu}a_2^{n_1-j}}{(a_1+a_2)^{n_1+n_2-\nu}}\mu^{j-1}\Big({j-1\atop\nu-1}\Big)\\
\fl=\frac{\Gamma(n_1+n_2-1)}{\Gamma(n_1)\Gamma(n_2)}\frac{a_1^{n_2-1}a_2^{n_1-j}}{(a_1+a_2)^{n_1+n_2-1}}\mu^{j-1}\,_2F_1\Big(1-j,1-n_2;2-n_1-n_2;\frac{a_1+a_2}{a_1}\Big).
\end{eqnarray}
The Gauss hypergeometric function in the second step terminates due to presence of negative parameters and equals the finite series appearing in the preceding step. Now, we plug this expression in (\ref{dwm}) and use the following summation result
\begin{equation}
\label{impsum}
\sum_{\mu=1}^{j-1}(-1)^\mu\left({j-1\atop \mu}\right)\mu^{j-1}=(-1)^{j-1}\Gamma(j).
\end{equation}
This can be proved by noting that $x^l$ is infinitely differentiable for non-negative integer $l$, and therefore if we consider $l=j-1$, then $d^{j-1}x^{j-1}/dx^{j-1}|_{x=0}=\Gamma(j)$. On the other hand, if we consider instead of $w(x)$, the function $x^{j-1}$ in either of the summations on two sides of (\ref{lrderiv}), then we obtain $d^{j-1}x^{j-1}/dx^{j-1}|_{x=0}=\sum_{\mu=1}^{j-1}(-1)^{\mu+j-1}\left({j-1\atop \mu}\right)\mu^{j-1}$. Therefore, we have the result (\ref{impsum}), and consequently from (\ref{dwm}),
\begin{eqnarray}
\label{left}
\nonumber
\fl\lim_{\epsilon\to0^{-}}\frac{1}{\epsilon^{j-1}}\sum_{\mu=0}^{j-1}(-1)^{\mu+j-1}\left({j-1 \atop \mu}\right)w(\mu\epsilon)\\
\fl=\frac{\Gamma(n_1+n_2-1)}{\Gamma(n_1)\Gamma(n_2)}\frac{a_1^{n_2-1}a_2^{n_1-j}}{(a_1+a_2)^{n_1+n_2-1}}\,_2F_1\Big(1-j,1-n_2;2-n_1-n_2;\frac{a_1+a_2}{a_1}\Big)
\end{eqnarray}

Similarly, it can be shown that the right $(j-1)$-th derivative is given by
\begin{eqnarray}
\label{right}
\nonumber
\fl\lim_{\epsilon\to0^{+}}\frac{1}{\epsilon^{j-1}}\sum_{\mu=0}^{j-1}(-1)^\mu\left({j-1 \atop \mu}\right)w((j-1-\mu)\epsilon)\\
\fl=\frac{\Gamma(n_1+n_2-1)}{\Gamma(n_1)\Gamma(n_2)}\frac{(-1)^{j-1}a_1^{n_2-j}a_2^{n_1-1}}{(a_1+a_2)^{n_1+n_2-1}}\,_2F_1\Big(1-j,1-n_1;2-n_1-n_2;\frac{a_1+a_2}{a_2}\Big).
\end{eqnarray}
Now, we can use the Pfaff transform result for nonnegative $l$,
\begin{equation}
_2F_1(-l,a;c;z)=(1-z)^l \,_2F_1\left(-l,c-a;c;\frac{z}{z-1}\right),
\end{equation}
provided all of $a,c,c-a$ are not integers in the interval $[1-l,0]$~\cite{V2007}. For (\ref{right}), this condition requires that $j-1<n_1+n_2-1$ and then the Pfaff transform can be applied, which transforms (\ref{right}) identically to (\ref{left}). In other words, the left and right derivatives agree for $j=2,3,...,n_1+n_2-1$. If this condition is not met, then the derivatives from two sides do not agree at $\lambda=0$. It can be also seen that (\ref{fm}) and (\ref{fp}) agree for the limit $\lambda\to 0$ when $j=2,3,...,n_1+n_2-1$.

%%%%%%%%%% REFERENCES %%%%%%%%%%

\end{document}